\documentclass[letter]{aa}  

\usepackage{graphicx}
\usepackage{hyperref}
\hypersetup{colorlinks, allcolors=blue}
\usepackage[varg]{txfonts}
\begin{document} 
    
    \titlerunning{}
    \authorrunning{Cameren Swiggum et al.}
    \title{The Radcliffe wave as the gas spine of the Orion arm}

    \author{C. Swiggum
          \inst{1},
          J. Alves\inst{1},
          E. D'Onghia\inst{2,3},
          R.A. Benjamin\inst{4},
          L. Thulasidharan\inst{3},
          C. Zucker\inst{5},
          E. Poggio\inst{6,7},
          R. Drimmel\inst{7},
          J.S. Gallagher III\inst{2},
          A. Goodman\inst{8}
          }

    \institute{
             Department of Astrophysics, University of Vienna, Türkenschanzstrasse 17, 1180 Wien, Austria\\
                \email{cameren.swiggum@univie.ac.at}
            \and 
                Department of Astronomy, University of Wisconsin-Madison, 475 North Charter Street, Madison, WI 53706, USA
            \and
                Department of Physics, University of Wisconsin-Madison, 1150 University Ave, Madison, WI 53706, US
            \and
                University of Wisconsin-Whitewater, Department of Physics, 800 West Main St, Whitewater, WI, 53190, USA
            \and
                Space Telescope Science Institute, 3700 San Martin Drive, Baltimore, MD 21218, USA
            \and
                Universit\'e C\`ote d'Azur, Observatoire de la C\`ote d'Azur, CNRS, Laboratoire Lagrange, Nice, France
            \and
                Osservatorio Astrofisico di Torino, Istituto Nazionale di Astrofisica (INAF), I-10025 Pino Torinese, Italy
            \and
                Center for Astrophysics $\vert$ Harvard $\&$ Smithsonian, 60 Garden St., Cambridge, MA 02138, US
             }

   \date{Received ...; accepted ...}

 
  \abstract
  {The Radcliffe wave is a $\sim3$ kpc long coherent gas structure containing most of the star-forming complexes near the Sun.  In this Letter we aim to find a Galactic context for the Radcliffe wave by looking into a possible relationship between the gas structure and the Orion (local) arm. We use catalogs of massive stars and young open clusters based on \textit{Gaia} Early Data Release 3 (EDR3) astrometry, in conjunction with kiloparsec-scale 3D dust maps, to investigate the Galactic \textit{XY} spatial distributions of gas and young stars. We find a quasi-parallel offset between the luminous blue stars and the Radcliffe wave, in that massive stars and clusters are found essentially inside and downstream from the Radcliffe wave.  We examine this offset in the context of color gradients observed in the spiral arms of external galaxies, where the interplay between density wave theory, spiral shocks, and triggered star formation has been used to interpret this particular arrangement of gas and dust as well as OB stars, and outline other potential explanations as well. We hypothesize that the Radcliffe wave constitutes the gas reservoir of the Orion (local) arm, and that it presents itself as a prime laboratory to study the interface between Galactic structure, the formation of molecular clouds in the Milky Way, and star formation.}


   \maketitle
%

\section{Introduction} 

The spiral arms of disk galaxies play an intrinsic role in the dynamics of their stellar and interstellar material and are even theorized to trigger the onset of star formation \citep{Roberts1969-jw}. There are a multitude of astrophysical sources observed to trace spiral structure including masers, OB stars, HII regions, interstellar dust, and Cepheids. Detailed images of face-on spiral galaxies reveal elongated dust lanes that often trace spiral patterns, but they are typically segmented and offset from the unobscured, blue-stellar component, which is an observation supported by the density wave theory of spiral arms \citep{lin_spiral_1964, shu_six_2016}. The arrangement of our own Galaxy's spiral arms is more ambiguous due to the Sun's embedded location near the Galactic midplane. The first detection of the Sun's nearest spiral arm (named the Orion arm by \cite{van_de_hulst_spiral_1954}\footnote{Also known as the local arm}) was made by \cite{Morgan1953-jg} using a limited sample of OB star parallaxes. Later studies of HII regions, 21-cm emission, parallax measurements of masers, and Cepheids have been interpreted in the context of a four-armed model of the Milky Way \citep{georgelin_spiral_1976, taylor_pulsar_1993, russeil_star-forming_2003, reid_trigonometric_2014, reid_trigonometric_2019, minniti_using_2021-1}, with the Orion (local) arm often absent or present only as a minor spur. 

The precise astrometry measured for over one billion stars by the ESA \textit{Gaia} mission \citep{gaia_collaboration_gaia_2021} has transformed our understanding of the Milky Way's structure and the ability to measure the distances and reddening for millions of stars near the Sun accurately has unveiled the local interstellar medium (ISM) in three dimensions \citep{Green2019-ew,Lallement2019-bn,lallement_updated_2022-1,Chen2019-qu,Rezaei_Kh2020-in,leike_resolving_2020}. Recently, \cite{Alves2020-um} discovered a $\sim3$ kpc long, narrow (aspect ratio of 1:20), and undulating gas structure they named the Radcliffe wave, connecting many nearby star-forming complexes (CMa, MonR2, Orion, Taurus, Perseus, Cepheus, North America nebula, and Cygnus) into a quasi-linear feature on the \textit{XY} plane of the Galaxy. The discovery of this gas structure was made possible by the compendium of accurate cloud distances \citep{Zucker2020-gj}, together with complementary distance measurements toward low column density clouds using the same method. When including ESA \textit{Gaia} data, the \cite{Zucker2020-gj} method reaches median distance errors to clouds of about 5\%, or about a factor of 5-6 times better than previously possible, transforming our perception of the gas distribution in the local Galaxy. 

The Radcliffe wave supersedes the long-held model of star formation in the solar neighborhood, namely the Gould Belt \citep{Gould1874-ed}, involving a presumed ring of molecular gas and OB associations tilted at a 20$^\circ$ angle from the Galactic plane \citep[e.g.,][]{Palous2016-rx}. The compendium of molecular cloud distances from \citet{Zucker2020-gj}, complemented with measurements toward lower column density clouds in \cite{Alves2020-um}, show that the "Gould Belt clouds" break down into two families of  star-forming molecular clouds belonging to two much larger structures, the Radcliffe wave and the split, which is an apparent spur-like feature situated between the Orion (local) and Sagittarius-Carina arms \citep{Lallement2019-bn}.\footnote{See interactive \href{https://faun.rc.fas.harvard.edu/czucker/Paper_Figures/radwave.html}{Figure 2} in \cite{Alves2020-um}.}

Two intriguing results emerge from the spatial structure of the Radcliffe wave: One result concerns its coherent undulation about the Galactic plane following a damped sinusoidal pattern, possibly originating from instabilities arising between the Milky Way's disk and halo \citep{Fleck2020-uo} or due to perturbations caused by the passage of a dwarf galaxy \citep{thulasidharan_evidence_2022}. The second result concerns the relation between the Radcliffe wave and the Orion (local) arm of the Galaxy. The Radcliffe wave appears to cross the Orion arm in the plane of the galaxy as traced by masers \citep{Reid2016-ht}, as seen in the interactive \href{https://faun.rc.fas.harvard.edu/czucker/Paper_Figures/radwave.html}{Figure 2} from \cite{Alves2020-um}, making it hard to assert if there is a relationship between the two structures. In this Letter, we focus on the latter of these two results. Specifically, we seek to understand if the Radcliffe wave is part of the Milky Way's spiral structure when considering tracers of the Orion (local) arm such as OBA-type stars, open clusters, and masers. 

Section 2 of this Letter outlines the data used for studying Galactic spiral structure. Section 3 describes our analysis and results. Section 4 attempts to contextualize our results with respect to density wave theory and makes connections to spiral structure as seen in some nearby external disk galaxies.

\section{Data}

\begin{figure*}[!hbt]
    \centering
    \includegraphics[width = \linewidth]{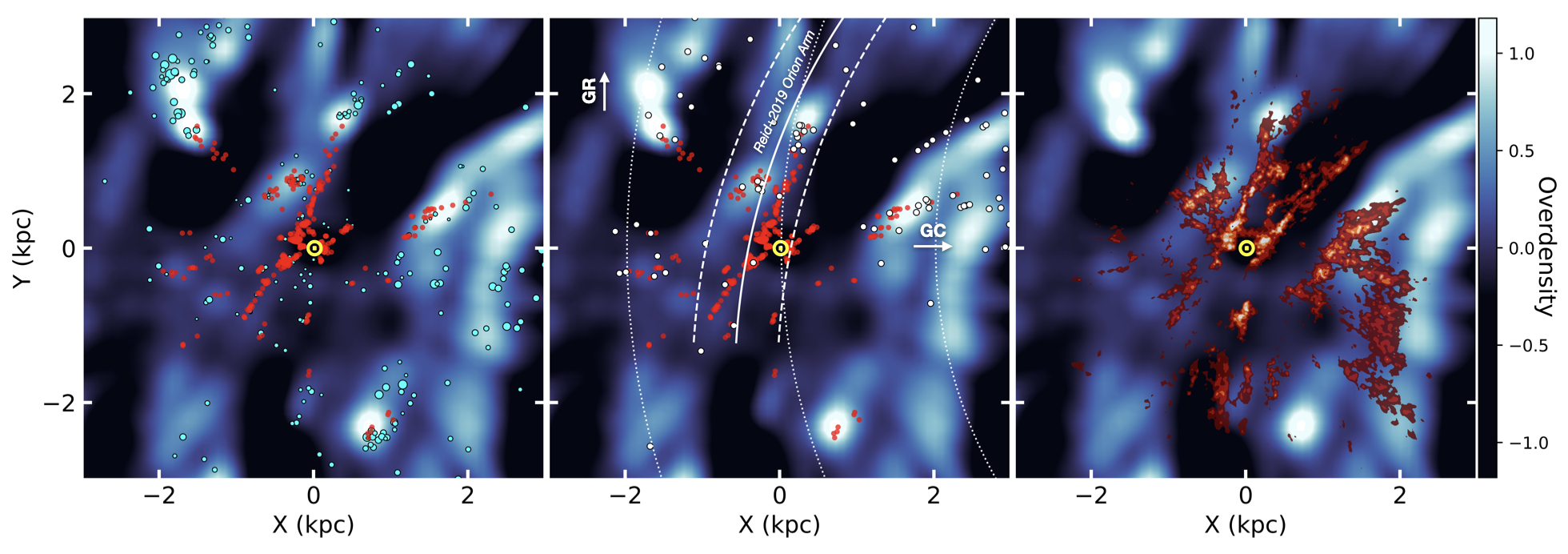}
    \caption{Galactic \textit{XY} (bird's-eye) view of various stellar and dust distributions within $\sim4.25$ kpc of the Sun (centered yellow dot). The over-density map of the OBA-type stars from \cite{Zari2021-dd} (in black, blue, and white) is shown in the background of each panel. The emergent spiral features in this map, from decreasing to increasing values in \textit{X}, are the Perseus, Orion, and Sagittarius-Carina arms. The star-forming clouds from \cite{Zucker2020-gj} are displayed as red points in the first two panels, with the Radcliffe wave appearing as a narrow alignment of clouds from $(X, Y) \approx (-1, -1)$ kpc to $(X, Y) \approx (0.5, 2)$ kpc. The cyan points in the first panel are the open clusters from \cite{Cantat-Gaudin2020-tk} with ages of less than 30 Myr and sized relative to the square root of their membership count. The second panel shows the masers (white points) and the fit to the Orion (local) arm (solid white line) from \citep{reid_trigonometric_2019}, along with the fit's width represented by the two dashed white lines. The second panel also displays three circles of constant, Galactocentric radius (dotted white lines) and two arrows pointing in the directions of the Galactic center and Galactic rotation. Overlaid in the third panel is the dust map of \cite{Lallement2019-bn}, with a heat map color-scale showing increasing dust density.}
    \label{fig:main_fig}
\end{figure*}

We use the Galactic \textit{XY} positions of molecular clouds from \cite{Zucker2020-gj}, complemented with distances to clouds in the Radcliffe wave from \cite{Alves2020-um}. The line-of-sight (LOS) distances to these clouds were computed using a Bayesian method with optical photometry from the Panoramic Survey Telescope and Rapid Response System \citep[PanSTARRS1;][]{flewelling_pan-starrs1_2020,Chambers2016-sq}, the National Optical Astronomy Observatory (NOAO) source catalog \citep{Nidever2018-uc}, near-infrared photometry from the Two Micron All Sky Survey \citep[2MASS;][]{Skrutskie2007-rn}, and parallaxes from the second \textit{Gaia} data release \citep[\textit{Gaia} DR2;][]{Brown2018-oh}.  

The availability of all-sky, multiband photometric surveys and precise parallaxes from \textit{Gaia} has led to the development of a multitude of 3D maps tracing interstellar dust at varying resolutions and spatial extents \cite[e.g.,][]{Green2019-ew,Lallement2019-bn,Chen2019-qu,Rezaei_Kh2020-in,leike_resolving_2020}. Dust mapping at kiloparsec scales has revealed traces of spiral structure, including the Orion (local) arm \citep{Chen2019-qu,Green2019-ew,Lallement2019-bn}. In this work we use the map of \cite{Lallement2019-bn}, which was constructed using a Bayesian inversion technique applied to \textit{Gaia} DR2 and 2MASS data. The resulting map has dimensions of $6\times6\times0.8$ kpc and achieves a 25 pc resolution for the portion of the map considered in this work.

Additionally, we investigate the Galactic \textit{XY} distribution of OBA-type stars from \cite{Zari2021-dd}. This all-sky sample was constructed with color-color and color-magnitude cuts using photometry and astrometry from \textit{Gaia} EDR3 and the Two Micron All Sky Survey \citep{Skrutskie2007-rn}. The sample was selected to be more complete, rather than pure; the authors indicated that some contamination might come from intermediate-mass and evolved high-mass stars. They used the EDR3 astrometry combined with expected proper motions from Galactic rotation to derive ``astro-kinematic'' distances and to remove spurious sources following this analysis. Additional spurious sources were accounted for using the classifier developed by \cite{rybizki_classifier_2022}, leading to a final sample count of 435,273 stars. \cite{poggio_galactic_2021} show that the Z21 data traces spiral structure, specifically the Sagittarius-Carina, Orion, and Perseus spiral arms, hence making the sample suitable for our study. Another tracer of young, bright, and massive stars are open clusters. We use 167 of the Gaia DR2 derived open clusters from \cite{Cantat-Gaudin2020-tk} with inferred ages of less than 30 Myr. Several works have recently shown that open clusters can trace nearby spiral structure \citep[e.g.,][]{Xu2018-um, poggio_galactic_2021,castro-ginard_milky_2021}. Masers also play an important role in mapping spiral arms at kiloparsec scales due to their intense luminosities and their location within massive star-forming regions. We also utilize the catalog of masers from \cite{reid_trigonometric_2019} and we especially consider the log-spiral fit to the subset of masers that have been identified as the Orion (local) arm.


\section{Methods and results}
\label{sec:methods_and_results}
\subsection{Dust-star offset}
\label{sec:offset}
\begin{figure}
    \centering
    \includegraphics[width=\linewidth]{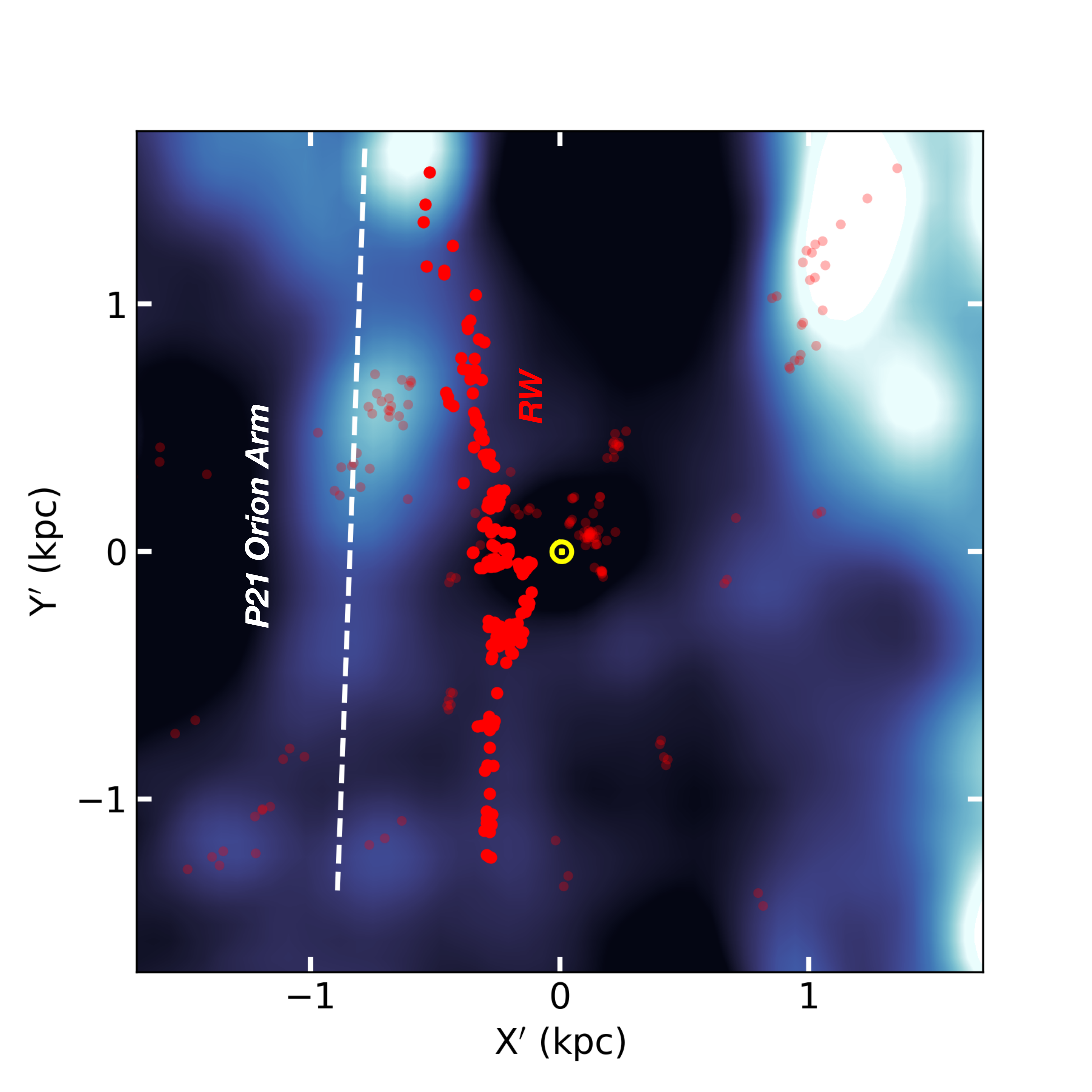}
    \caption{Zoomed and rotated version of Figure \ref{fig:main_fig} displaying the Z21 stellar over- and under-density (in black, blue, and white) and  star-forming region locations \citep{Zucker2020-gj} in red, with the bold-red points marking clouds belonging to the Radcliffe wave. The dashed white line shows the fit to the over-density peaks along the P21 Orion arm. The Sun is located at the center in yellow.}
    \label{fig:oba_peaks}
\end{figure}

We replicated the method of \cite{poggio_galactic_2021} (hereafter P21) to plot the Z21 sample density in Galactic \textit{XY} coordinates: Two bivariate Epanechnikov kernels were used, one with a bandwidth of 0.3 kpc to measure the local density and the other with a bandwidth of 2 kpc to measure the mean density. These density maps were input into Equation (1) of P21, resulting in the clear spiral structure seen in Figure \ref{fig:main_fig}. P21 originally applied this method to their upper main sequence (UMS) sample of stars, but additionally applied it to the Z21 sample (see their Figure B.1), finding that both samples, in a consistent way, recover traces of the Sagittarius-Carina, Perseus, and Orion arms (hereafter the P21 Orion arm). 

Figure \ref{fig:main_fig} also displays the Galactic \textit{XY} positions of the star-forming clouds from \cite{Zucker2020-gj}, with the Radcliffe wave apparent as a thin stretch of clouds from $(X, Y) \approx (-1, -1)$ kpc to $(X, Y) \approx (0.5, 2)$ kpc. There is a clear offset between the OBA-type stellar component of the Orion arm and the dust composing the Radcliffe wave. To determine the size of this offset, we first rotated the OBA density counterclockwise 32 degrees, such that the P21 Orion arm was oriented vertically in a new coordinate frame $XY^\prime$. We then smoothed the collapsed, 1D profiles along $X^\prime$ by $5\sigma$ in order to trace the average shape of the P21 Orion arm more easily, rather than its individual over-density clumps. We located the smoothed profile peaks using the SciPy peak-finding algorithm\footnote{scipy.signal.find\_peaks} and then fit them with a line, which we consider to be the "spine" of the P21 Orion arm in the section we are considering (Figure \ref{fig:oba_peaks}). The resulting offset between the P21 Orion arm spine and their nearest point along the Radcliffe wave ranges between 250 to 730 pc with a median of 580 pc. 

The Radcliffe wave is also apparent as an elongated dust filament in the 3D map of \cite{Lallement2019-bn} (Panel 3 of Fig. \ref{fig:main_fig}) and agrees with \cite{Zucker2020-gj} distance estimates to molecular clouds, hence the dust-star offset is also seen in the diffuse dust. The authors in Z21 note a spatial mismatch between their hot-luminous star sample and the dust. However, since Z21 did not initially find a clear spiral structure in their sample, this extended linear offset between the Radcliffe wave and the P21 Orion arm was not obvious.

While both young open clusters and masers are shown to trace spiral structure in the Milky Way \citep{reid_trigonometric_2019, castro-ginard_milky_2021}, their spatial distributions in the context of the aforementioned dust-star offset are more ambiguous (Figure \ref{fig:main_fig}). The spatial distribution of young open clusters (Fig. \ref{fig:main_fig}; panel 1) coincides with the locations of the P21 spiral arms; however, the number of OCs is too sparse to unveil a general offset from the dust of the Radcliffe wave alone. A mismatch between the fit to the masers of the Orion arm from \cite{reid_trigonometric_2019} and the Radcliffe wave was first mentioned by \cite{Alves2020-um}. This mismatch is again shown in this work (Figure \ref{fig:main_fig}, panel 2); the Radcliffe wave has a local pitch angle of $\sim29$ degrees and the P21 Orion arm has a similar pitch angle of $\sim31$ degrees \citep[see Figure 5 of][]{poggio_galactic_2021} for a more extended view. This angle is in disagreement with the results of \cite{reid_trigonometric_2019} who report an Orion (local) arm pitch angle of 11.4 degrees. 

The Orion (local) arm fit from \cite{reid_trigonometric_2019} yields a width of $310\pm{50}$ pc, meaning that despite the discrepancy in pitch angle, the narrow Radcliffe wave is contained within the broad distribution of the Orion (local) arm masers. These masers are generally located in regions of higher OBA-type star overdensity along the P21 Orion arm (Figure \ref{fig:main_fig}; middle panel). It is noted by P21 that the OBA-type stars and masers tend to overlap in the first and second Galactic quadrants for all three arms recovered by the OBA-type stars, but a lack of maser observations in the third and fourth quadrants renders comparisons in these directions difficult. Additionally, parallax measurements from masers, while precise at distances exceeding Gaia’s capabilities, are lacking in quantity compared to our large sample of stellar parallaxes and molecular cloud distances, giving us a more precise view of the Orion arm at distances within 2 kpc from the Sun, whereas the masers are capable of tracing the Orion arm out $4.7$ kpc from the Sun in the first quadrant \citep[see Figure 1 of][]{reid_trigonometric_2019}. Classical Cepheid samples also probe spiral structure at distances exceeding this work; however, their spatial distributions do not show clear signs of a present Orion arm \citep{gaia_2022}. Further work is warranted in order to untangle the varying spatial distributions of different Galactic spiral arm tracers.

An important issue is the role of the star formation region that we label as "Cepheus Far" (Figure \ref{fig:labeled_radwave}). This active star-forming complex – which contains, among other members, the HII regions Sh2-140, Sh2-150, four masers, and the OB association Cep OB3 -- lies within the P21 Orion arm.  \cite{reid_trigonometric_2019} identified this as an Orion (local) arm star-forming region. Recently, \cite{pantaleonigonzalez_alma_2021} have revived the possibility, first suggested by \cite{Morgan1953-jg}, that this star-forming region is part of a ``branch'' of the Orion arm which they name the ``Cepheus Spur.'' This spur is traced by the OB associations Cep OB4, Cam OB1, Aur OB1, and Gem OB1. Given the intersection of this structure with the P21 Orion arm, further investigation of this claim is warranted. 


\subsection{Stellar mass estimate}
\label{mass}
We compared the gas mass of the Radcliffe wave to the young stellar mass for a portion of the P21 Orion arm. To select the segment of stars along the arm that are parallel with the Radcliffe wave, we drew a cylinder around the P21 Orion arm spine (Figure \ref{fig:oba_peaks}) which selects 18575 stars within a 400 pc radius, which was chosen visually to match the width of the arm. The invoked lower mass limit of this stellar subsample is, in principle, that of an A0V-type star ($\sim2\text{M}_{\odot}$); however, this value is rather uncertain since it is shown by Z21 that there exists a contamination of lower mass F-type stars which may also be present as unresolved binaries. Taking into account an extreme case where half of the Z21 sample has a spectral type later than an A0V-type star, we constructed a \cite{Kroupa2001-ah} initial mass function (IMF)\footnote{https://github.com/keflavich/imf} and varied the initial mass until the number of stars above 2.0$\text{M}_{\odot}$ was in the range of 0.5 to 1.0 times the number of stars in the selected sample. The total initial mass of the simulated stellar population is then $ 1.4~\text{--}~2.8 \times 10^{5} M_{\odot}$ and the Radcliffe wave mass was calculated to be $\sim3\times10^{6}~\text{M}_{\odot}$ based on column density measurements provided by the Plank dust map \citep{planck_collaboration_planck_2014, Alves2020-um}. Assuming a star formation efficiency range of $1-3~\%$ for the gas of the Radcliffe wave, the initial stellar mass of the P21 Orion arm is $2-10$ times the fraction of which the Radcliffe wave could have produced, implying that the stars of the P21 Orion arm formed from previous gas structures.

\section{Discussion}

The combined discoveries of spiral structure in the young-luminous stars \citep{poggio_galactic_2021} and the unique linear arrangement of major star-forming regions composing the Radcliffe wave \citep{Alves2020-um} present a fascinating picture of the nearest spiral arm to the Sun. The offset found between the two young structures in this work confirms that spiral arms are density waves \citep{lin_spiral_1964, toomre_group_1969, Toomre1977-vp, shu_six_2016}.

While it is established that spiral arms are density waves, significant debate remains about their lifetimes. Numerical experiments suggest that spiral arms are transient features, \citep{dobbs_spurs_2006, Foyle2011-vv, sellwood_lifetimes_2010, wada_interplay_2011, roskar_radial_2012, Kawata2014-al, perez-villegas_galactic_2015, sellwood_discriminating_2019}, or that they fluctuate locally in density but are statistically long-lived \citep{fujii_dynamics_2011, DOnghia2013-cv}. It is important to note that multiple observational works have supported long-lived spiral arms via a variety of different tracers \citep{donner_structure_1994, zhang_secular_1998, Martinez-garcia_spiral_2009, Martinez-garcia_signatures_2013, scarano_radial_2013}. While this work does not prove that spiral density waves trigger star formation, our resulting view of the Orion arm complements previous observations suggesting that they can, at least, have sufficient longevity to configure gas into kiloparsec-scale gas spines \citep{Elmegreen1986-ad}.

Our mass estimate of the P21 component directly offset from the Radcliffe wave (Section \ref{mass}) reveals that the molecular gas at present is not sufficiently massive to have produced the offset stellar mass. This result supports an evolutionary scenario for the Orion arm in which previous gas segments have passed through an over-density, shocked, and collapsed to form the visible young and bright stars. The excess gas, which failed to collapse and form stars eventually, has dissolved, with perhaps some residual gas remaining in the P21 Orion arm (Figures \ref{fig:main_fig}, \ref{fig:labeled_radwave}). The Radcliffe wave is then the present-day gas spine of star-forming regions and will also dissolve, similar to the gas spines that previously produced the young stars downstream from it. This scenario also requires the consideration of Galactic inflows as the source of sustained star formation in the Orion arm, as the gaseous disk would need to be replenished over time based on the mass ratio between the Radcliffe wave and the P21 Orion arm. In this picture, the Radcliffe wave would be the elongated filament of gas collected at the spiral structure over-density, indicating that all of the star-forming regions along the wave have originated as a consequence of a Galactic spiral shock.

\begin{figure}[ht!]
    \centering
    \includegraphics[width=\linewidth]{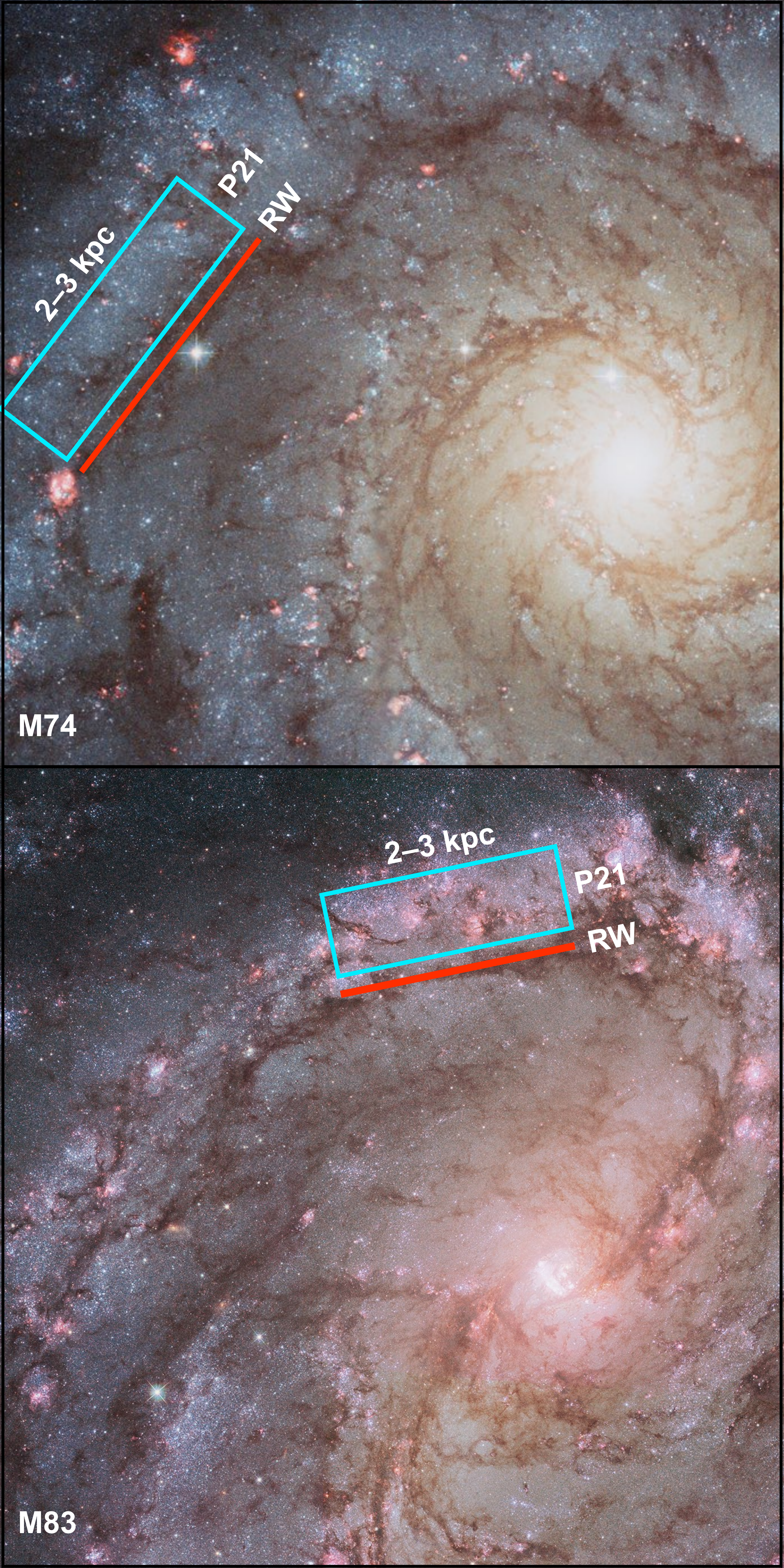}
    \caption{Two Hubble Space Telescope images, with the top panel showing the grand-design spiral galaxy Messier 74 (M74) which demonstrates a prominent dust lane and a luminous, blue stellar region, offset from each other in one of its spiral arms. The bottom panel displays a more flocculant spiral galaxy, Messier 83 (M83), which tends toward a disordered configuration of dust, gas, and stars which may be more analogous to the Milky Way. The overlaid drawings are meant to represent the resemblance of the Orion arm's dust and OBA offset shown in Figures \ref{fig:main_fig} and \ref{fig:oba_peaks} to those seen in these face-on spiral galaxies, with the red line and the blue rectangle depicting the Radcliffe wave and the P21 Orion arm, respectively. These representations of the Orion arm have been drawn approximately to scale based on the distance measurements to M74 and M83.}
    \label{fig:extra_gal}
\end{figure}

Recent works based on \textit{Gaia} hint at the location of a surface mass over-density responsible for the arrangement of the Radcliffe wave and the P21 Orion arm. The study of \cite{eilers_strength_2020} used \textit{Gaia} DR2 astrometry along with other all-sky surveys to study the kinematics of luminous red giant branch (RGB) stars, assuming a steady-state model of a logarithmic spiral arm. They used the radial component (in the direction of the Galactic center) of the stars' velocities and found two nonaxisymmetric features which they could directly relate to stellar surface mass density via models (Figures 3, 4; \cite{eilers_strength_2020}). One of these features is near the Orion (local) arm; a visual inspection reveals that the Radcliffe wave partially overlaps with this over-density and lies, at most, 1 kpc downstream from it. In the steady-state density wave theory \citep{lin_spiral_1964}, this feature is expected within the corotation radius since the Radcliffe wave and the emerging young stars rotate faster than the pattern speed of the density wave and eventually overtake it. However, outside of the corotation radius, the pattern speed of a spiral arm is faster than Galactic rotation, and hence the Radcliffe wave would be expected to lie directly upstream from the over-density instead, and the P21 Orion arm would instead lie upstream from the wave. There are some caveats with this model that complicate the interpretation:

\begin{enumerate}
    \item{Spiral arms are not in a steady state but constantly changing in density and they are not always logarithmic.}
    \item{The Radcliffe
wave shows a vertical sinusoidal undulation (Alves 2020), which is not expected in the density wave theory \citep[see][]{thulasidharan_evidence_2022}.}
    \item{The steady-state model of logarithmic spiral applies better to grand-designed spiral galaxies. It is important to note that this is not the case of the Milky Way, which seems to be multiarmed with more than one corotation radius.}
\end{enumerate}

Furthermore, the average age of the P21 Orion arm is unknown due to the uncertainty of its stellar composition (the fraction of O-, B-, and A-type stars present). In the Lin-Shu density wave theory, it is expected for age gradients traced by intrinsic stellar color gradients to be present across luminous stellar arms. From the Radcliffe wave to the downstream edge of the P21 Orion arm, there exists a gradient in the stars' \textit{Gaia} BP-RP colors (blue to red); however, this is likely a product of dust reddening. There is no obvious color gradient across the P21 Orion arm alone, nor is there an obvious age gradient in the open clusters overlapping in this region, but this requires additional analysis incorporating multiband photometry and spectroscopy. Additionally, if the P21 Orion arm has an average age of 100 Myr (assuming a sample of mostly A-type stars), with its spatial offset from the upstream \cite{eilers_strength_2020} feature being 1-2 kpc, then the expected rotational velocity difference between the two structures would be in the range of 10-20 kms$^{-1}$. This may become observable with the expanded radial velocity catalog of \textit{Gaia} DR3.



While a large fraction of local star-forming clouds lie along the Radcliffe wave, there are also clouds that make up the "split" \citep[Figure \ref{fig:main_fig};][]{Lallement2019-bn, Zucker2020-gj} such as Sco-Cen, Aquila, and Serpens. It is not clear if the split is related to the Orion arm; however, its linearity and its direct overlap with the same feature from \cite{eilers_strength_2020} indicate that this structure and its star-forming regions might be part of the same global picture. Solidifying the formation history of the Radcliffe wave and the split will require future 3D kinematic analysis, which will become increasingly viable with the future data releases from \textit{Gaia} and other ground-based spectroscopic surveys measuring radial velocities.

The damped, sinusoidal shape of the Radcliffe wave is still a complete mystery. Photometric studies of edge-on disk galaxies have revealed bending waves with amplitudes and wavelengths similar to the Radcliffe wave \citep{narayan_wobbly_2020}, suggesting that such corrugations might be common in the dust lanes of spiral arms. If the Radcliffe wave's undulation is somehow related to gas instabilities \citep{Fleck2020-uo}, stellar feedback, or perturbations in the Galactic potential \citep{Antoja2018-nr}, one would expect to find more global corrugations in the dust and young stars across the Milky Way. Such structures have been investigated in OB stars \citep{pantaleonigonzalez_alma_2021} and signatures of varying oscillation amplitudes across stars of different ages (5-100 Myr) are present across the disk \citep{thulasidharan_evidence_2022}. Global corrugations in the dust map of \cite{lallement_updated_2022-1} have also been found, including regions along the Orion arm in the second quadrant and in the Sagittarius-Carina arm in the first and fourth quadrants.

In summary, the formation histories of the Radcliffe wave and the offset P21 Orion arm remain unsolved; however, we find that their spatial arrangement poses a striking resemblance to the spiral arms of many external disk galaxies. Messier 51 (M51) is a remarkable example of a grand design spiral galaxy exhibiting extended dust lanes lined along the edges of its luminous blue spiral arms. An analysis of the density waves produced by older stars in M51 exhibits explicit offsets from the dust lanes in one of its arms \citep{Egusa2017-ux}. In the top panel of Figure \ref{fig:extra_gal}, we display a Hubble Space Telescope image of another grand design spiral galaxy, Messier 74 (M74), which exhibits extended dust lanes offset from its blue spirals, making a clear connection to the spatial arrangement of the Radcliffe wave and the P21 Orion arm (Figures \ref{fig:main_fig} and \ref{fig:oba_peaks}) with similar spatial scales based on the distance to M74. As previously mentioned, the Milky Way is likely not a grand design spiral galaxy \citep{castro-ginard_milky_2021} and even the existence of coherent structure in the prominent Perseus arm has been recently questioned \citep{peek_burtons_2022}. The bottom panel of \ref{fig:extra_gal} shows M83, which is an example spiral galaxy with more disordered spiral arms, yet still exhibiting extended dust lanes offset from unobsurced, luminous stellar arms, indicating that it is not quite a flocculent galaxy. The ability to detect distinct, parallel, and linear structures of the young stars and the dust suggests that the Milky Way is not a flocculent spiral, and instead, similar to M83, it lies somewhere in between a galaxy with unambiguous spiral structure and one in which spiral arms are less traceable due to disordered dust and stars.

The spatial picture presented in this work further demonstrates that the star-forming regions composing the Radcliffe wave (Figure \ref{fig:labeled_radwave}) have a connected history and indicates that the wave is an important laboratory for star formation, not only on the scales of its individual complexes, but also on the scale of a Galactic spiral arm. Future work will employ a dynamical analysis incorporating the expanded radial velocity catalog of \textit{Gaia} DR3 in order to pinpoint the physical processes that established this arrangement of dust and luminous stars.

\section{Acknowledgments}
This work has made use of data from the European Space Agency (ESA) mission Gaia \url{https://www.cosmos.esa.int/gaia}, processed by the Gaia Data Processing and Analysis Consortium (DPAC, \url{https://www.cosmos.esa.int/web/gaia/dpac/consortium}). Funding for the DPAC has been provided by national institutions, in particular the institutions participating in the Gaia Multilateral Agreement.

\bibliographystyle{aa}
\bibliography{full_bib}

\begin{appendix}
\section{Labeled star-forming regions of the Radcliffe wave}
\begin{figure}[hbt!]
    \centering
    \includegraphics[width=\linewidth]{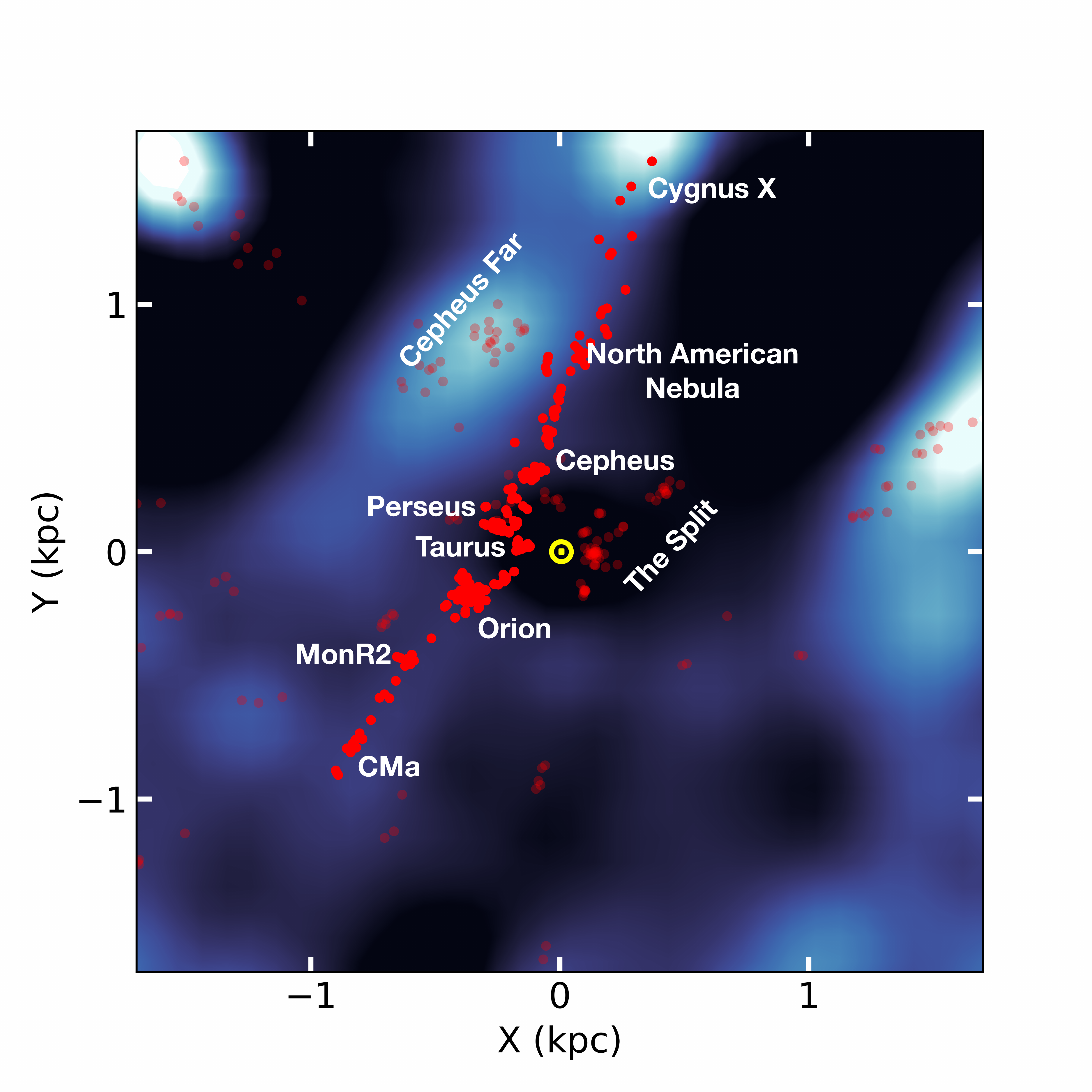}
    \caption{Zoomed version of Figure \ref{fig:main_fig} displaying the Z21 stellar density (in black, blue, and white) and the Radcliffe wave clouds \citep{Zucker2020-gj} in red, with labels placed next to its major star-forming regions. Both the "split" and "Cepheus Far," which are not part of the Radcliffe wave, are labeled as well. The Sun's location is marked in yellow.}
    \label{fig:labeled_radwave}
\end{figure}
\end{appendix}


\end{document}